# Interstellar Probes: The Benefits to Astronomy and Astrophysics

### K. F. Long[a]*

[a] *Initiative for Interstellar Studies, The Bone Mill, New Street, Charfield, GL12 8ES, United Kingdom*,
kelvin.long@i4is.org
* Corresponding Author

**Abstract**

Long range observations in the field of astronomy have opened up our understanding of the Solar System, the Galaxy and the wider Universe. In this paper we discuss the idea of direct in-situ reconnaissance of nearby stellar systems, using robotic probes. In particular, we consider what additional knowledge can be learned that can only be obtained by such close encounters. This may include calibration of existing measurements, detailed observations of stellar winds, astrometry measurements of stellar parallax, refinement of our understanding of physics through the use of long baseline interferometers. In addition, getting close to an exoplanet will enable detailed knowledge of planetary interiors, surface processes, geological evolution, atmospheric composition and climate, internal seismology, detailed surface morphology and even the speculative possibility of detecting the presence of microbial life, detailed palaeontology or even indigenous life-forms. We argue that astronomical remote sensing should be pursued in parallel with in-situ reconnaissance missions by robotic probes, so that both can enhance the discoveries and performance of the other. This work is in support of Project Starshot; an effort to send a Gram-scale probe towards another star at 0.2c within the next two decades, and return images and other data to the Earth. Presented at the 47$^{th}$ IAA Symposium on Future Space Astronomy and Solar-System Science Missions. Session on Space Agency Strategies and Plans.

**Keywords:** Interstellar Probes, Astronomy, Astrophysics,

## 1. Introduction

In this paper we explore the benefits to astronomy and astrophysics of sending a space probe towards the exoplanet of another star. We do not argue here that one is better than the other, but both are needed and complimentary as will be shown later in this paper. However, it is also a fact that there are certain science investigations that can only be achieved by close proximity.

Space telescopes have seen remarkable success, building on from the ground based telescopes. The Hubble Space Telescope has bought about a sea change in our visions of the universe. The Kepler Space Telescope has opened our eyes to the wider plethora of planetary worlds that appear to be around all stars.

Some of these missions have also enabled a different perspective on the planet Earth. One of the first views of Earth at the outset of the 'space age' was taken by Project Apollo 8 in 1968 with a famous photo known as 'Earthrise', which appears to show the Earth rising above a lunar crater. Voyager 1 went better than this when it took the 'Pale Blue Dot' image in 1990, apparently showing the Earth suspended in a sunbeam, from a distance of 6 billion km. Recently in 2013 the Cassini mission took 'the Day the Earth smiled' photo from a distance of 1.4 billion km. It showed the Earth-Moon system hovering remotely under the rings of Saturn. These images have a transformative potential on the human consciousness, and one can only speculate how the first close-up views of a planet around other stars will inspire people in the arts and sciences.

Space probes that explore our solar system have also seen great success. Two recent space missions worth highlighting includes the Cassini-Huygens to the ringed planet Saturn and its many moons, and the New Horizons mission to the dwarf planet Pluto. The scientific community awaits with great anticipation the results of the JUNO mission which arrived at Jupiter in July 2016.

Our efforts to go further, outside of the Solar System, are sadly lacking however. The Voyager probes stand as the single greatest achievement. Launched in 1977, both probes are now at a distance of 143 AU (Voyager 1) and 118 AU (Voyager 2) [1], where 1 AU = $1.496 \times 10^{11}$ m, or the mean Earth-Sun distance. They would take tens of thousands of years to reach their nearest line of site stars at their current speeds of 17 km/s. It is estimated that the power supply on both spacecraft will run out around the year 2025.

Any spacecraft that goes outside of the Solar System and beyond will need to be equipped with high performance and reliable technologies. This will include Radioisotope Thermoelectric Generators (RTGs), low and high gain antenna's, thrusters for manoeuvring, radiators for heat transfer, protection shields to mitigate impacts from interstellar dust and high energy ions, efficient computer storage systems, star tracker navigation systems. This is in addition to the suite of scientific instruments, such as magnetometer booms, cosmic ray detectors, particle flux detectors, optical cameras, spectrometers, an on-board telescope to name just a few. A proper discussion of these technologies and their application to a deep space mission is beyond the scope of this paper.

There are ideas to send probes further afield. Missions that go beyond the Voyagers and towards the Oort cloud are known as interstellar precursor missions. One such example was the 1,000 Astronomical Unit probe study from JPL in the 1990s [2]. Another example was a 200 AU study performed by the International Academy of Astronautics [3] for which this author was a contributor. Currently the Johns Hopkins Applied Physics Laboratory is looking at a potential 1,000 AU mission to be sent by the year 2032 travelling at 20 AU/year it would arrive half a century later [4].

Missions that go beyond the Oort cloud into true interstellar space and towards the nearest stars are true interstellar missions. In the 1970s members of the *British Interplanetary Society* designed Project Daedalus, a 450 tons payload that would be carried by a fusion powered spacecraft, carrying 50,000 tons of Deuterium-Helium-3 [5].The flyby spacecraft would travel to the nearest stars travelling at 36,000 km/s or 0.12c, reaching its stellar target within half a century and passing through without decelerating. Although this is considered a landmark study in interstellar spacecraft design, the projections for its likely realization placed it two centuries ahead, mainly due to the need for an independent off-world based economy to fund such a large scale endeavor.

Recently, another effort to revisit Project Daedalus is under way, called Project Icarus [6]. It seeks to re-design Daedalus using our improved knowledge of science and technology. It also aims for full deceleration into the target system instead of just a flyby mission, and it would take around a century to get there.

Another study conducted in the 1980s set out to design an interstellar mission architecture that did not carry its own fuel. This resulted in the Starwisp concept [7, 8]. This was a 1 ton spacecraft sent to the stars at 34,000 km/s or 0.11c pushed by a 65 GW microwave beam. This would be placed in orbit, in addition to a 560,000 tons Fresnel lens to collimate the diverging beam for the maintenance of a consistent pressure profile across any sail surface.

These projects just illustrate some of the good efforts to attempt design solutions for future spacecraft missions. But they all involve large masses, highly challenging mission architectures, and massive costs, which push them into the far future of possibilities, rather than the near-term.

However, another project that is currently underway, which builds on some of this historical work, is the Breakthrough Initiatives Project Starshot [9]. The project was launched in April 2016. It aims to address the issues of cost being a strong function of the spacecraft mass, but also that carrying fuel is a limiter on performance capability due to the nature of the ideal rocket equation.

The Breakthrough Starshot solution, is a mission to send a Gram-scale probe to the stars at 60,000 km/s or 0.2c within 20 years, using a ground based 100 GW laser, transmitted through the atmosphere, to push a 4 m orbiting solar sail in under ten minutes acceleration. Many technology and physics obstacles have been identified to make this mission possible, and an initial seed fund of $100 million has been sponsored to facilitate fundamental Research and Development. Although there are many technical problems which must be solved before Starshot can be realized, the possibility that something similar to the Starshot architecture will be sent in the coming decades is high.

The above provides the context where missions to the stars in future decades are becoming a real possibility and it is not just science fiction. Given this, it is useful to begin to ask what the science case is to justify the cost of such missions, as opposed to building long-range observatories.

A real world example of where this can be seen very clearly is when comparing the image taken by the Hubble Space Telescope of the dwarf planet Pluto in 2003 to that taken by New Horizons in 2015 (see Fig. 1). The amount of surface detail observed is overwhelmingly superior from New Horizons.

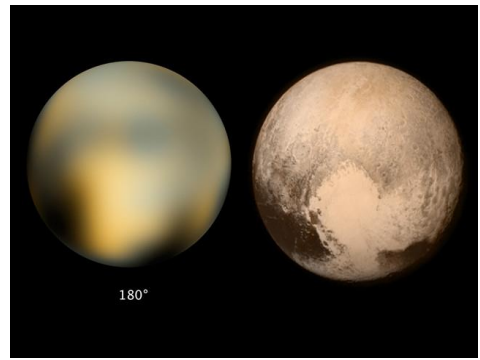

Fig. 1. Pluto as seen by the Hubble Space Telescope and the New Horizons spacecraft (credit: NASA/Johns Hopkins University Applied Physics Laboratory [10])

**2. Astronomical Targets for Future Missions**

Within the Stellar neighbourhood of the Sun, there are multiple stars of different spectral types, mass, luminosity, age and distance contained in over 20 different systems [11]. These stars include a mix of binary systems, triple star systems, known exoplanetary systems, and those with dust disks. The stars vary from Sun-like to red dwarfs to white dwarfs and even brown dwarfs. The decision on which star system to send a space probe to will depend on three elements. That is (1) Convenience (2) Astrophysics (3) Astrobiology. Let us consider each of these in turn.

*2.1 Convenience*

When we say convenient we are really talking about the ability to build something and also whether the economic cost of launching such a mission is realistic. A reasonable distance in which we may in the coming century be able to send a small reconnaissance space probe, would be to around 20 LY (1 LY = $9.4607 \times 10^{15}$ m). Even if we could travel at the speed of the Breakthrough Starshot probe, around 0.2c, this mission would still take a century. This is likely therefore the outer bounds for any space missions for which we should give our priority.

We could send probes further, but this requires a different level of autonomous capability and may imply a requirement for artificial intelligence. The probability of mission failure for a longer mission also increases, as components fail year on year. The necessity for self-repair may also then push up the payload mass.

The further the distance the more difficult the antenna transmitter-receiver issues become and our ability to detect any weak signals, distinguished from the background noise of the natural astrophysical sources.

Another consideration which relates to the speed of the spacecraft is the transit time of the encounter. If we assume that the size of a stellar system, out to its debris cloud is given as a function of the stellar mass $m$ relative to the mass of our sun $m_s$, then the radius of the system will be around $\sim 0.1(m/m_s)^{1/3}$ [12]. Given this, it can easily be shown that for Proxima Centauri for example, the debris radius would be around 0.05 LY, and the encounter time at Daedalus speed of 0.122 would be 295 days and at Starshot speed of 0.2c it would be 179 days. This is how much time any spacecraft would have to use its instrumentation and gather as much scientific data on the system as possible before it exited the system. For the Starshot mission, with an estimated cost for the full flight of around $10 billion, this amounts to ~$56 million per day of in-situ observations at the target. This has neglected the possibility of conducting observations en-route, and of course once the infrastructure is in place many more of these small probes can be launched.

Within 20 LY the nearest stars that we may choose to send a mission too which is most convenient includes the Alpha/Beta Centauri system (4.3 LY) and its faint companion Proxima Centauri (4.2 LY), Barnard's star (5.9 LY), and Wolf 359 (7.9 LY).

*2.1 Astrophysics*

This refers to the types of physics we may be interested in learning about, from the stars or the surrounding disk material or any exo-planets in orbit. The type of stars could also vary in type, from main sequence to flare stars or brown dwarf, or binary systems.

The Alpha/Beta Centauri system and its companion Proxima is a good candidate due to the three different star types, namely; Yellow star, Orange star, and a red dwarf. This is also a good system to study because the stars are similar in age to the Sun, and so comparable chromospheric activity, astroseismic studies and stellar rotations gives us insights into the structure and evolution of our own star.

There are also other star systems which may be of interest, including Groombridge 34 (11.7 LY) which exhibits random variations in its luminosity due to flares and so is considered a variable star. Another possibility might be Struve 2398 (11.5 LY) which is a binary star system, and both stars exhibit a type of variability common to flare stars. They also exhibit significant x-ray emission. Another possibility might be EZ Aquarii (11.3 LY) which is a triple star system where all three components are M-type red dwarfs.

Any decision on which stars may have the most astrophysical interest depends on the strategies of our scientific programs, and also the uncertainties in our modelling for which in-situ information would provide critical value.

*2.1 Astrobiology*

Given that no biology has yet been discovered outside of the Earth's biosphere, the decision on which stellar targets may characterise the greatest astrobiology interest depends on our assumptions about life (see section 4).

Conventionally, astrobiologists talk about a 'Goldilocks Zone' also known as a circumstellar habitable zone. The assumption is that any planet within this zone from its parent star would have planetary surface conditions to support liquid water at atmospheric pressure. Too close to the Sun and the radiant energy falling on the planet could be too high to allow for life's survival or even emergence. Too far from the Sun and the radiant energy could be too low, leading to too cold a condition for life.

The system that has the most interest currently is the star Proxima Centauri, where an Earth-like mass (27% more massive) has been located in the habitable zone of its parent star.

One potentially interesting star is Tau Ceti (11.9 LY). The star is similar to our sun in spectral type, exhibits little variation in its luminosity and so appears stable, and five exo-planets have been discovered there, of which two are thought to be in the habitable zone.

Exoplanets have also been found around Ross 128 b (11.03 LY) which is a red dwarf star. It is the second closest exoplanet known to date. The exoplanet is believed to be similar in mass to the Earth (35% more massive) and is located within the habitable zone of its parents star. A triple planetary system has also been discovered around the star Wolf 1061 (13.8 LY).

It is clear, that we are at the outset of what can be detected, since the first ever confirmed detection of an exo-planet in 1992, but the future looks promising and exciting. These exoplanets are discovered using a variety of methods, but this includes the transit method, radial velocity method, microlensing, pulsar timing Transit Timing Variation and direct imaging.

**3. Scientific Investigations**

There are many areas of science for which a deep space mission could add value and this is discussed in depth by Crawford [13, 14, 15]. Firstly, we must not forget that there is also the space between the stars which holds large scientific interest. This includes investigations of the Oort cloud, the pure interstellar medium and solar heliosphere, the potential discovery of new dwarf planets, brown dwarfs or even free floating planets.

There is also the potential of utilising the gravitational lensing point which is located between 550 – 1,000 AU. This is a result of the bending of light around the Sun due to gravity and was predicted by the General Theory of Relativity. Such a mission to this location, if launched in the opposite direction to Proxima Centauri, would be able to look back on it with a magnification of approximately $\sim 10^4$ times better than any observing platform in Earth orbit [16].

Any spacecraft launched towards the nearest stars could also obtain improved calibration of existing measurements and properties of those stars. It could also conduct detailed observations of the stellar atmosphere and stellar winds.

Another great benefit of such a mission would be to conduct astrometry measurements for stellar parallax. Accurately knowing the distance to a star is important for correctly determining its astrophysical properties. Sending a space probe towards another star will provide an opportunity to make parallax measurements with a baseline over multiple light years distance. This is approximately 10,000s of times longer than present methods which use the semi-major axis of the Earth's orbit as a baseline. Currently, parallax measurements are accurate to about 20 pc (1 parsec = 3.26 LY) from the Sun. A longer baseline would allow accurate measurements of stellar distances of more than 1.2 million pc, allowing accurate determination of the distances of trillions of stars.

Any close-up observations of exoplanets orbiting around distant stars would enable investigations of the planetary interiors, surface processes and an improved understanding of their geological evolution.

A spacecraft that is travelling at interstellar distance will likely be moving fast. With this comes the opportunity to test refinements of Special and General Relativity Theory but also to investigation other speculative ideas such as related to breakthrough propulsion physics. It could also conduct observations to help our understanding of dark matter and dark energy.

Sending a space probe towards another star will also allow significant information on the properties of any exoplanets within the system. In our search for planets around other stars we have discovered Hot Jupiter's, Super Earth's, Tidally-locked planets and they range in compositions from mostly iron to mostly water.

Our current knowledge of exoplanets is limited to their orbital semi-major axes, orbital period, eccentricity, mass, radius, and an inferred bulk density. Some analysis can be performed on the atmospheric composition using spectroscopy. Currently we can only guess as to how diverse the climates, geological processes and surface morphologies of these different worlds will be.

If a space probe were to visit one of these systems it would give additional knowledge of the local moons of those planets, knowledge of any ring structures. A close-examination of the surfaces would also allow a detailed knowledge of the geological processes and surface morphology at work. We could also gain insights into the internal planetary structure. If we were able to penetrate the atmosphere with a sub-probe we could study the atmospheric composition and climate. Placing any landers onto the surface would give us knowledge of seismology, local gravitational and magnetic fields, radiogenic isotope dating in rock samples.

**4. Life in the Universe**

If we were able to place a spacecraft on to the surface of an exoplanet it may be possible to search for the presence of microbial life-forms or even indigenous life-forms. Evidence of a separate bio-genesis event would have profound implications for the understanding of the place of Homo sapiens in the universe.

Even if we arrive on another planet and find that any higher life-forms have gone extinct and only microbial life remains, our ability to conduct detailed investigations through a form of palaeontology would teach us a lot about how biological organisms grow in different planetary biospheres under evolution by natural selection. Such investigations would likely need planetary landers craft on the ground.

In terms of looking for life on other planets, there are five types of categories that we might consider.

*Type 1*: These are planets which appear to have uninhabitable surfaces but might support a sub-surface biosphere.

*Type 2*: These are planets which appear habitable such as spectroscopic evidence of water and carbon dioxide.

Type 3: These are planets for which plausible atmospheric bio-signatures are detected.

*Type 4*: These are planets which appear habitable but also show emissions consistent with our expectations for low level industrialisation (e.g. Pollutants in the atmosphere or chemical depletion of an ozone layer).

*Type 5*: These are planets which have the elements of the other categories but also show strong evidence of the occupation by advanced intelligence due to its activities within its system (e.g. Dyson Spheres).

The detection of industrialization on any scale around another planet is termed 'techno-signatures' and in terms of priorities for any future missions this is likely to get our most interest.

Most of the focus of the above discussion has been on the life that we know and our assumptions about carbon-based chemistry. Our best understanding to date is that life (that is animals, plants) is distinguished from inorganic matter by homeostasis – a property of a system such as the concentration of a substance in solution that is actively regulated to remain near constant. For example, for mammals like us, this could be through body temperature, the pH level of the extracellular fluids, or the concentration of Sodium, Potassium, Calcium ions and glucose in the blood plasma. We then define life as being composed of cells, which undergo metabolism, can grow, adapt to their environment, respond to stimuli and reproduce.

However, in our quest to understand the nature of intelligence in the universe, we have to at least admit the possibility that 'life' or 'living systems' [17] may be characterised by different combinations of chemistry or even by non-chemical processes.

In 1944 the physicist Erwin Schrödinger wrote "*living matter, while not eluding the laws of physics as established up to date, is likely to involve other laws of physics hitherto unknown which however once they have been revealed will form just as integral a part of science as the former….life can be defined by the process of resisting the decay to thermodynamic equilibrium*" [18].

To illustrate three examples of systems we might study that could exhibit complex behaviour, in a method that is analogous neuron functioning in a brain, but instead as a kind of networked intelligence, here are three potential ideas:

*Idea 1 Space Plasmas*. A plasma is typically blown off of a star from a stellar wind. It consists of ions and electrons, bound together by electromagnetic fields. For a cloud of plasma that is drifting in deep space for millions of years, provided there is some means of occasional energy transfer through the system, is it possible for some level of self-organization to occur such that it is analogous to the 'black cloud' [19] in the famous story by Fred Hoyle?

*Idea 2 Mycelium fungus*. This is a bacterial colony consisting of a mass of branching hyphae and is typically found in soils where it absorbs nutrients from their environment by the secretion of enzymes onto a food source and then breaking down biological polymers into smaller units called monomers. This process is vital for the decomposition of organic material. Is it possible that some material like mycelium could evolve to some level of networked intelligence if it grew to a large enough scale [20]?

*Idea 3 Conscious Stars*. The American physicist Greg Matloff has highlighted the interesting observation that cooler, less massive, redder stars in our stellar neighbourhood revolve around the centre of the Milky Way galaxy faster than their hotter, more massive and bluer stars. This is known as Parenago's discontinuity. Matloff has suggested that quantum mechanical effects may lend themselves towards a volitional star hypothesis [21].

It is not the purpose of this paper to argue for the credibility of these ideas, but just to illustrate the nature of living systems that may not meet our accepted definitions. These three ideas are just examples of what are currently not on the radar for any future space missions, since they would struggle to simultaneously meet our accepted definitions of 'life' and 'intelligence'. This author suggests that in a universe with a large variety of types of stars and planets, that it may also be possible for there to be a wide variety of intelligent systems.

## 5. Discussion

The different types of probes we could send towards the system of another star will vary in type. The simplest type of mission is a flyby. The next would be to conduct a flyby, but in some way slow the vehicle a little for the purposes of increased encounter time which required full deceleration. Then there is full orbital insertion into the target system. But given that stars move with a proper motion of order 100-200 km/s and any probe sent to the star with decades travelling time would have to be travelling at 10,000 km/s, there is a large difference in the velocity gradient to be overcome to make this achievable.

However, if it was possible to decelerate then this also opens up the possibility of deploying orbital spacecraft around the star and planets, or even atmospheric penetrators or impactors. Then if this was possible so too may be landers so that in-situ samples could be taken. Data would then be relayed back to any orbiting satellites and then back to the main spacecraft for transmission back to Earth.

There is little possibility of sample return to Earth, since the mission duration is then doubled and this also presents significant technology and system architecture issues (e.g. local fuel acquisition).

Finally in this section, it would be useful to discuss the argument of 'telescope versus probe'.

If the Breakthrough Initiatives Project Starshot is successful in sending a Gram-scale probe to the nearest stars within two decades, it would reach its mission target a further two decades later. By the time the data got back to Earth it would be approaching half a century. Given this, it is likely that parallel technology developments on Earth would have resulted in massive improvements in Earth based or space orbit telescope systems. One is then forced to confront how a single image of an exoplanet, taken from a passing interstellar probe, can be justified, given a long-range telescope of the future may be able to achieve similar levels of data acquisition, but presumably for reduced cost?

We can imagine a scenario with an on-board spacecraft telescope with an aperture $D_p$, positioned at a set distance $d_p$ from its target star which is a distance of $d_s$ from sol. This can be compared to the required diameter $D_s$ of a solar system based optical interferometer, which is related by $D_p d_s/d_p$ [22].

If we assume a 500 cm aperture space telescope positioned at 1 AU from a target star of Proxima Centauri at 4.2 LY distance to get the same optical performance a solar system interferometer would have to be 1,328 km in size, which is ~38% the lunar diameter.

This is illustrated in Table 1-4, which is calculated assuming a Proxima Centauri mission target at 4.2 LY assuming on-board telescope apertures of 500 cm (Daedalus-like [5]), 100 cm, 10 cm and 1 cm (Starshot-like [9]).

Table 1. Projection of Solar System based interferometer requirements (km) for space telescope apertures (cm) at a set distance from target star (AU).

|  | 500 cm | 100 cm | 10 cm | 1 cm |
|---|---|---|---|---|
| 0.1 AU | 13,280 | 2,656 | 265.6 | 26.56 |
| 1 AU | 1,328 | 265.6 | 26.56 | 2.656 |
| 10 AU | 132.8 | 26.56 | 2.656 | 0.265 |
| 100 AU | 13.28 | 2.656 | 0.266 | 0.027 |
| 1000 AU | 1.328 | 0.266 | 0.027 | 0.003 |

* AU = Astronomical Unit

Breakthrough Starshot depends very much on the progress in electronic miniaturisation in addition to the nanotechnology of materials science. It also depends on a projected reducing costs of near-infrared lasers (i.e. $100/W down to 5 cents/W), and increasing cost of laser power (i.e. kW up to GW required). Assuming the spacecraft could reach its stated target of the Alpha Centauri system, located 4.3 LY distance it would then have the incredibly difficult challenge of beaming data back using a power source that is of order 20 W. A large receiver (much bigger than the current Deep Space Network) back on Earth would then have to try and pick this up. This may require something similar to the 1970s NASA Cyclops study [23]. This same small spacecraft would need to carry adequate scientific instruments to conduct actual science measurements that are useful.

Although Breakthrough Starshot is likely to cost a lot to launch the first mission (i.e. $10 billion current estimate), once the systems architecture is in place, it has the luxury that it can send many swarms of low mass probes which presents the possibility of both instrumentation variety, but also networked data systems for information consolidation. The integrated data from a large number of probes may build up to an equivalent information value to data from a larger mass single probe.

Clearly, to justify any such mission, science measurements that go above and beyond what can be achieved by remote observations, must be a critical element of the mission. In this paper, we have laid out the broad scope of areas for which scientific surveys should be focussed.

Finally, some words on the potential scalability of this technology. This is on the assumption that some form of laser/microwave beaming capability was present in the future. Under such an assumption, it may be possible to beam energy direct to satellites to provide them with power, or indeed to beam power to distant spacecraft in the Solar System. It may also be possible to supply energy to the national grid of planet Earth, from a space orbital position.

In terms of deep space missions, if the laser power could be scaled high enough, this even opens up possibilities for propelling human crews to the distant planets and the nearest stars. Such a concept was discussed by Jones [24] who suggested that a 500 person crew in a 2.2 million tons vessel, could be accelerated to 0.1c at an acceleration of 1 m/s$^2$, using a sail diameter of 6,000 km and would require a total power of 1,200,000 TW. This sounds like a vast number, but it just shows the long-term potential of such technology. If a crew could be placed in orbit around a distant star or orbiting exoplanets, the amount of science that can be done is likely to be exceptional. Although it is accepted that such missions cannot currently be considered to be near-term.

**6. Conclusions**

In this paper we have considered the scientific benefits of sending a probe towards another stellar system. In particular we have highlighted the benefits to astronomy and astrophysics.

The prediction of this author is that for a single Starshot probe arriving at its system four decades from now, we will find it difficult for this to compete with advanced astronomical platforms based around Earth, in terms of mission cost and scientific value. However, once the ability to send multiple probes is in place, a point will be reached where the value and cost of the space probes will be unquestionable and difficult to compete with. It would be useful to conduct a trade-study to assess say how many Starshot probes would be equivalent to an advanced Earth orbiting Space Telescope performance half a century from now.

Yet with both long-range observatories and in-situ space probes, they should be seen as parallel and necessary programs which supplement and compliment the observations of each other. Only when both are widely applied, across the electromagnetic spectrum, gravity waves and other domains of reality we may choose to investigate, can we truly begin to understand our place in the Universe.

**Acknowledgements**

The author would like to thanks Ian Crawford for discussions, as well as John Davies and Rob Swinney for comments on an earlier version of this paper.